\documentclass[nohyperref]{article}

\usepackage{microtype}
\usepackage{graphicx}
\usepackage{subfigure}
\usepackage{booktabs} 

\usepackage{hyperref}

\usepackage[accepted]{icml2022}

\usepackage{amsmath}
\usepackage{amssymb}
\usepackage{mathtools}
\usepackage{amsthm}

\usepackage{mathtools}

\DeclarePairedDelimiterX{\infdivx}[2]{(}{)}{%
  #1\;\delimsize\|\;#2%
}

\usepackage[capitalize,noabbrev]{cleveref}

\theoremstyle{plain}

\theoremstyle{definition}

\theoremstyle{remark}

\usepackage[textsize=small]{todonotes}

\icmltitlerunning{Bayesian Light Source Separator}

\begin{document}

\twocolumn[
\icmltitle{Scalable Bayesian Inference for Detection and Deblending in\\Astronomical Images}



\icmlsetsymbol{equal}{*}

\begin{icmlauthorlist}
\icmlauthor{Derek Hansen}{equal,umstats}
\icmlauthor{Ismael Mendoza}{equal,umphys}\\
\icmlauthor{Runjing Liu}{voleon}
\icmlauthor{Ziteng Pang}{umstats}
\icmlauthor{Zhe Zhao}{umstats}
\icmlauthor{Camille Avestruz}{umphys}
\icmlauthor{Jeffrey Regier}{umstats}
\end{icmlauthorlist}

\icmlaffiliation{umstats}{Department of Statistics, University of Michigan}
\icmlaffiliation{umphys}{Department of Physics, University of Michigan}
\icmlaffiliation{voleon}{The Voleon Group}

\icmlcorrespondingauthor{Ismael Mendoza}{imendoza@umich.edu}
\icmlkeywords{astronomy, catalog, deblending, Bayesian}

\vskip 0.3in
]



\printAffiliationsAndNotice{\icmlEqualContribution} 


\section{Introduction}
The forthcoming generation of astronomical surveys will peer deeper into space, revealing many more astronomical light sources than their predecessors. Because of the greater density of light sources in these surveys' images, many more light sources will visually overlap.
Visually overlapping light sources, called ``blends'', are expected to make up 62\% of the galaxies imaged by the upcoming Legacy Survey of Space and Time~\cite{sanchez2021effects}.
Blends are challenging for traditional (non-probabilistic) astronomical image processing pipelines because they introduce ambiguity into the interpretation of the image data.

We present a new probabilistic method for detecting, deblending, and cataloging astronomical objects called the Bayesian Light Source Separator (BLISS). BLISS is based on deep generative models, which embed neural networks within a Bayesian model and use deep learning to facilitate posterior inference.
BLISS generalizes StarNet~\cite{liu2021variational}, which can only analyze images of starfields.

In the BLISS statistical model (\cref{model}), the latent space is interpretable: one random variable encodes the number of stars and galaxies imaged, a random vector encodes the locations and fluxes of these astronomical objects, and another random vector encodes the galaxy morphologies.
Conditional on these random variables, the data (i.e., the pixel intensities in a collection of astronomical images) are modeled as Poisson or Gaussian.
Owing to this Bayesian formulation, BLISS requires no special logic to analyze blended galaxies.

For posterior inference (\cref{inference}), BLISS uses a new form of variational inference based on stochastic optimization, deep neural networks, and the forward Kullback-Leibler divergence.
This new methodology is known as ``Forward Amortized Variational Inference'' (FAVI)~\cite{ambrogioni2019forward}. In FAVI, a deep encoder network is trained on data simulated according to the generative model to solve the inverse problem: predicting the latent variables that generated a particular synthetic astronomical image. FAVI has scaling advantages over Markov chain Monte Carlo and achieves improved fidelity of the posterior approximation compared with traditional variational inference in our application.

Algorithmically, inferences in BLISS are produced by a sequence of three deep convolutional encoder networks, each conditioned on the output of the earlier network (\cref{amortization}). The first encoder performs detection, estimating the number of light sources in particular regions. Conditional on samples from the first encoder, the second encoder probabilistically classifies each sampled source as a star or a galaxy. Conditional on sampling a galaxy by the second encoder, a third encoder network estimates the shape/morphology. No restrictive assumptions are made about the factorization of the posterior approximation, as is common in more traditional approaches to variational inference.

The BLISS inference routine is fast, requiring a single forward pass of the encoder networks on a GPU once the encoder networks are trained.
BLISS can perform fully Bayesian inference on megapixel images in seconds, and produces more accurate catalogs than traditional methods do (\cref{experiments}).
BLISS is highly extensible, and has the potential to directly answer downstream scientific questions in addition to producing probabilistic catalogs (\cref{conclusion}).

\section{The Statistical Model}
\label{model}
Our generative model consists of two parts: the prior distribution over all possible astronomical catalogs and the likelihood of an image given a particular catalog.

\subsection{Prior}

Let $\mathcal{Z}$ be the collection of all possible catalogs, and let $z \in \mathcal{Z}$ be a particular realization.
Our prior over $\mathcal{Z}$ is a marked spatial Poisson process.
Light sources arrive according to a homogeneous Poisson process with rate $\mu$, which is set based on prior knowledge.
In other words, for a given image of size $H \times W$, the number of sources $S$ in this image follows the Poisson distribution:
\begin{align}
	S &\sim \text{Poisson}(\mu HW).
	\label{eq:n_prior}
\end{align}
The locations of each of the sources $s = 1, \dots, S$ are uniform in the image:
\begin{align}
  \ell_s \mid S & \sim \text{Uniform}([0, H] \times [0, W]).
\end{align}
Each source $s$ is either a star or galaxy:
\begin{align}
  a_s \mid S \sim \text{Bernoulli}(\xi),
\end{align}
where $\xi$ is the proportion of imaged sources that are expected to be stars according to prior knowledge.


Given the point spread function (PSF) at a particular location in the image, a star's appearance is fully characterized by its flux $f_{s}$, which follows a truncated power law distribution:
\begin{align}
    f_{s, 1} \mid S, a_s = 1 & \sim \text{Pareto}(f_{min}, \alpha)
    \label{eq:flux_prior}.
\end{align}

Unlike stars, the shape of galaxies can vary greatly within an image.
To flexibly model this variety of shapes, we use a low-dimensional embedding following an uninformative prior to encode a galaxy's flux and shape: 
\begin{align}
v_s \mid S, a_s = 0 \sim \mathrm{Normal}(0, I_{D \times D}),
\end{align}
where $D$ is a user-defined embedding dimension.

\subsection{Likelihood}
\label{likelihood}

Let $z_s = \{\ell_{s}, a_{s}, f_{s}, v_{s}\}$ denote the latent variables describing light source $s$.
Let $z = \{S, \{z_{s}\}_{s = 1}^S\}$ be a catalog sampled from the prior.
Radiation from the light sources in catalog $z$ is recorded as the observed photoelectron count $x_{n}$ at each pixel $n$ of the astronomical image.
The photoelectron count $x_{n}$ follows a Poisson distribution with rate $\lambda_{n}(z) + \gamma_{n}$, where $\lambda_{n} (z)$ is a deterministic function of the catalog and $\gamma_{n}$ is background intensity.
In practice, since the number of arrivals is large, we use a normal approximation to the Poisson distribution:
\begin{align}
  x_n \mid z \sim \mathrm{Normal}(\lambda_n(z) + \gamma_{n}, \lambda_n(z) + \gamma_{n}).
\end{align}

The contribution of light source $s$ to the intensity of pixel $n$, denoted $\lambda_{n} (z_s)$, depends on whether source $s$ is a star or a galaxy.
If source $s$ is a star, the PSF and flux give its intensity at pixel $n$.
If source $s$ is a galaxy, then its contribution to pixel $n$ comes from a decoder neural network $g_{\theta} (z_{s})$, which is trained according to the procedure in \cref{sec:training}.



\section{Variational Inference}
\label{inference}

We infer the posterior distribution of the catalog, $p(z \mid x)$, by using variational inference (VI), which allows for computationally efficient approximate inference~\cite{blei2017variational, zhang2018advances}.
Rather than drawing samples like MCMC, VI turns the problem of posterior inference into a numerical optimization problem.
From a family of tractable distributions $q_\phi$, parameterized by $\phi \in \Phi$, VI aims to find the approximating distribution $q_{\phi^{*}}$ that minimizes a divergence metric to the posterior distribution.


Our generative model (\cref{model}) is \textit{transdimensional}: the number of light sources is not fixed within a given image.
Transdimensional inference can be challenging.
To allow for a variable number of light sources while maintaining tractability, we divide the image into $4 \times 4$-pixel subimages, which we call ``tiles.''.
For each tile $t = 1, \dots, T$, let $z_t$ represent the subset of the cataloged light sources with centers in tile $t$.
Our variational distribution factorizes across tiles and limits the number of sources centered in each tile $S_t$.
We further factorize the distribution for each tile into factors for each source indexed by $s^\prime = 1, \dots, S_t$.
These factors approximate the posterior densities of the location $\ell_{t, s^\prime}$,
source type $a_{t, s^\prime}$,
star flux $f_{t, s^\prime}$ 
and galaxy shape $v_{t, s^\prime}$.
\subsection{Amortization and model architecture}
\label{amortization}

\begin{figure*}
  \centering
  \includegraphics[width=\linewidth]{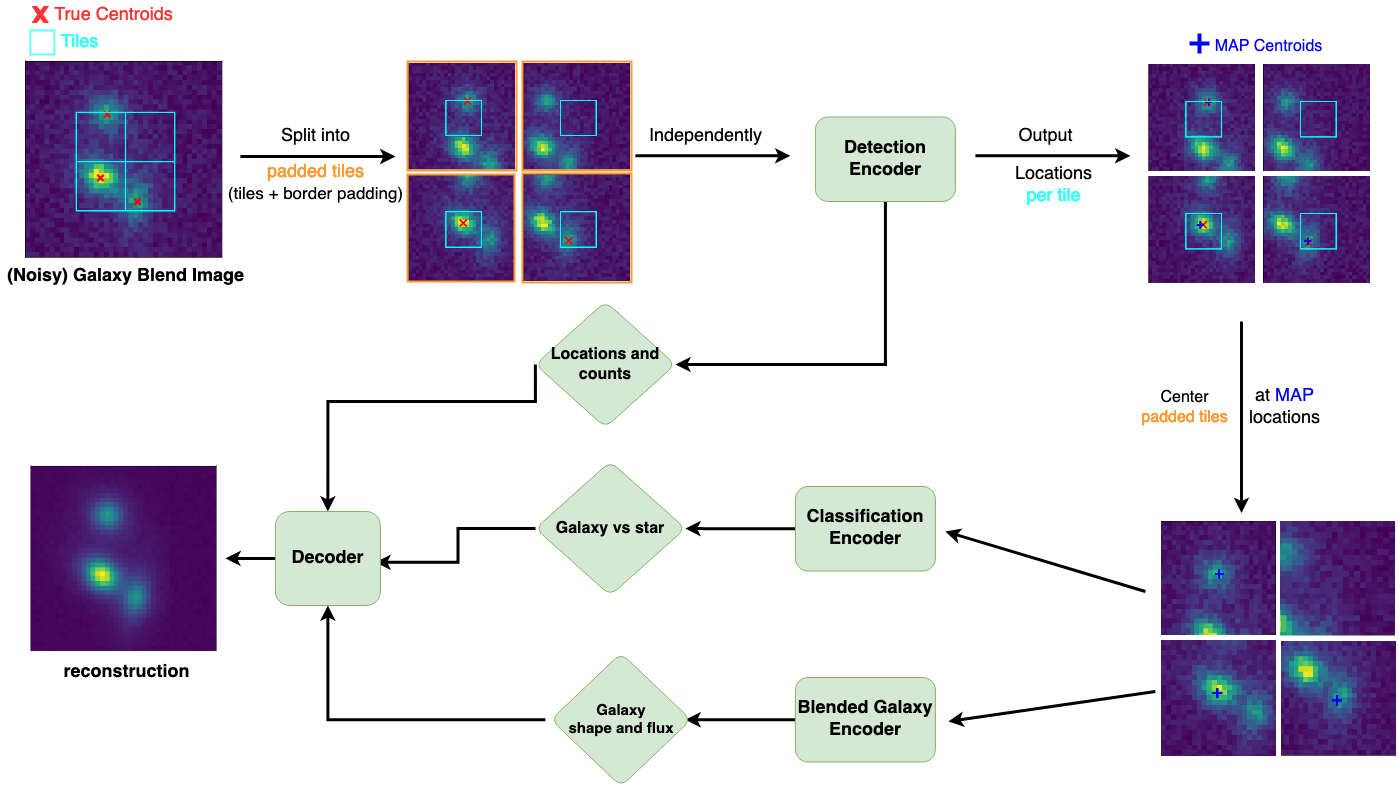}
  \caption{
    The BLISS encoder sequence.
  }\label{fig:tiling}
\end{figure*}
With traditional VI, fitting a variational distribution would require running a computationally intensive iterative optimization procedure once for each tile.
Instead, BLISS utilizes \textit{amortized} variational inference \citep{kingma2014auto, zhang2018advances}.
For each tile $t=1,\ldots,T$, we let $x_t$ denote the corresponding \textit{padded tile}: the $52 \times 52$-pixel cropped sub-image centered around the $4 \times 4$-pixel tile with index $t$.
In our amortized inference procedure, an encoder neural network is trained to transform each padded tile $x_t$ into distributional parameters $z_t$ for that tile.
After a training phase, amortized variational inference enables fast inference on unseen data; the average cost of training per-data point is reduced as more data is processed.
Also, amortized inference allows for stochastic optimization with subsamples of the image data;
during each optimization iteration, it is far more efficient to load a small part of the image and to compute gradients with respect to it than do so for the entire image.

The BLISS encoder consists of a sequence of neural networks that operate on padded tiles (\cref{fig:tiling}).
The padded tile $x_t$ is first fed to the \textit{detection encoder},
which infers the number of sources in each tile $S_t$ and the locations of each source $\ell_{t, s^\prime}$ for $s^\prime = 1, \dots, S_t$.
After sampling locations for each source, each padded tile is centered and fed to the \textit{classification encoder}, which infers whether any light source in the corresponding tile is a star or a galaxy.
Finally, fluxes $f_{t, s^\prime}$ are sampled for sources that are stars, and galaxy morphology parameters are sampled from the \textit{galaxy encoder}.

\subsection{Training procedure}\label{sec:training}
BLISS has two training stages.
First, we learn a tractable generative model of single galaxy shapes by fitting a variational autoencoder (VAE) \cite{kingma2014auto} to simulations from GalSim \cite{rowe2015galsim}.
Second, we train each of the aforementioned encoder networks (i.e., the {location encoder}, the {classification encoder}, and the {galaxy encoder}) on simulated data sampled from the generative model.

\paragraph{Galaxy VAE} We use a variational autoencoder to learn a low-dimensional representation $v \in \mathbb R^D$ of centered galaxies.
To generate galaxies, we place a prior on the flux, ellipticity, and size of the bulge and disk components as well as the angle of rotation.
These parameters are rendered into single, centered galaxies using the GalSim simulation package.
Separate encoder $g_\phi$ and decoder $f_\theta$ networks are trained to minimize the ELBO.
The learned decoder network $g_\theta$ is subsequently included as a component of our overall generative model.

\paragraph{Encoder networks}
The location and classification encoders are trained using forward amortized variational inference (FAVI) \citep{ambrogioni2019forward}.
FAVI uses the expected forward KL divergence as its training objective.
This has several advantages.
First, it leads to a better optimization path than the traditional VI objective, and it lets us avoid using the high-variance REINFORCE gradient estimator \citep{liu2021variational}.
Second, it correctly estimates of the marginal posterior distribution of latent variables, implicitly integrating over nuisance parameters such as the latent properties of light sources below the detection threshold \citep{ambrogioni2019forward}.


\section{Experiments}
\label{experiments}

To illustrate the performance of BLISS, we run the trained encoder
network on a $1489 \times 2048$ frame (run 94, camcol 1, field 12) from stripe 82 of the Sloan Digital Sky Survey (SDSS).
After training for 5.5 hours with synthetic data (a one-time upfront training cost), BLISS inferred a probabilistic catalog for this SDSS frame in just 10.5 seconds.

\begin{table}
    \centering
  \begin{tabular}{rcllll}
    && \multicolumn{2}{c}{BLISS} & \multicolumn{2}{c}{PHOTO} \\
    \cmidrule(l){3-4}
    \cmidrule(l){5-6}
    Mag & Ground Truth & TP & FP & TP & FP \\
    \midrule
  17 - 18 & 31 & 31 & 1 & 28 & 0 \\
18 - 19 & 39 & 38 & 2 & 37 & 2 \\
19 - 20 & 64 & 59 & 9 & 55 & 12 \\
20 - 21 & 117 & 111 & 20 & 104 & 20 \\
21 - 22 & 232 & 187 & 45 & 185 & 44 \\
22 - 23 & 386 & 126 & 91 & 117 & 47 \\
Overall & 889 & 571 & 175 & 545 & 130 \\

  \end{tabular}
\caption{Catalog comparison of BLISS to PHOTO, treating COADD as ground truth. Each row is a particular bin of galaxies based on their magnitude according to COADD. ``TP'' refers to true positives and ``FP'' refers to false positives.}\label{tbl:sdss_detections}
\end{table}

Although BLISS is probabilistic and outputs a distribution of catalogs, we use the mode of the variational distribution as a point estimate for comparison to non-probabilistic catalogs.
We let the coadd catalog from SDSS (henceforth, COADD) for this frame serve as a proxy for ground truth.
COADD based its estimates on all the filter bands of numerous frames, whereas BLISS used just the r-band of one frame; we expected COADD to serve as reasonable, though imperfect, proxy for the ground truth in our benchmarks.
This approach to benchmarking was developed in \citet{regier2019approximate}, where it is described in great depth.
We compare BLISS with PHOTO~\cite{lupton2005sdss}, which utilizes the same SDSS frame as BLISS, but has access to all five filter bands.

\cref{tbl:sdss_detections} compares the detection accuracy of BLISS and PHOTO.
To qualify as a match, a source must be within one pixel of the other in $L^{\infty}$ distance.
Overall, BLISS detects 571 out of the 889 total sources in the SDSS frame.
The vast majority of sources that BLISS did not find were faint (magnitude $\ge 21$), making them harder to detect.
Of the 7 sources greater than 20 magnitude that were not matched by BLISS, 4 sources were missed due to errors in COADD, 2 sources had unusual shapes, and 1 source was in a particularly difficult blend with many other sources.
This was determined by manually checking discrepancies with the more recent DECaLS survey \cite{dey2019overview}.
Similarly, of the 12 sources that BLISS detected that were unmatched in COADD, 10 were due to errors in COADD, and 2 were mistakes by BLISS.
These latter two mistakes were both cases where a source's center straddled adjacent tiles, leading to a prediction in both.
These types of errors are quite rare, and could be fixed by a post-processing step that conditions on neighboring tiles, or by a more complex variational approximation.

\begin{table}
    \centering
  \begin{tabular}{rllllll}
      &\multicolumn{3}{c}{BLISS} & \multicolumn{3}{c}{PHOTO} \\
    \cmidrule(l){2-4}
    \cmidrule(l){5-7}
       Mag & Tot & Gal & Star & Tot & Gal & Star\\
    \midrule
  17 - 18 & 0.97 & 1.00 & 0.96 & 0.96 & 1.00 & 0.95 \\
18 - 19 & 1.00 & 1.00 & 1.00 & 1.00 & 1.00 & 1.00 \\
19 - 20 & 0.98 & 1.00 & 0.96 & 1.00 & 1.00 & 1.00 \\
20 - 21 & 0.94 & 0.91 & 0.98 & 0.90 & 0.86 & 0.98 \\
21 - 22 & 0.81 & 0.86 & 0.73 & 0.83 & 0.81 & 0.87 \\
22 - 23 & 0.63 & 0.74 & 0.50 & 0.66 & 0.70 & 0.63 \\
Overall & 0.84 & 0.87 & 0.80 & 0.85 & 0.84 & 0.86 \\

  \end{tabular}
\caption{Accuracy of classifications made by BLISS and PHOTO.}
\label{tbl:sdss_accuracy}
\end{table}

\begin{figure}[!h]
    \centering
    \includegraphics[width=0.55\linewidth]{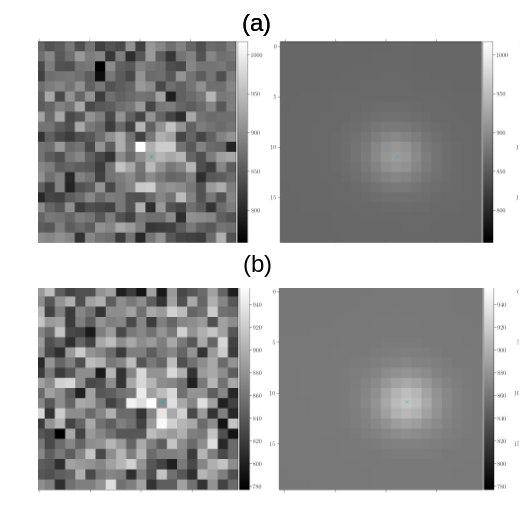}
    \caption{Images of BLISS detections near the 50\% detection threshold in SDSS images (left) and their reconstructions (right). Example (a) was detected in COADD. (b) was not present in COADD but found in the DECaLS catalog.}
    \label{fig:borderline}
\end{figure}

\cref{tbl:sdss_accuracy} compares the source-type classification accuracy of BLISS and PHOTO. Overall, BLISS correctly classifies most sources.
When sources are dimmer and more ambiguous, BLISS sensibly defers to the prior, labeling most of them as galaxies.
In SDSS images without much blending, BLISS identifies more sources than PHOTO, 
despite having less access to less data (only one of five bands), with the advantage of producing calibrated uncertainties for every prediction.
While BLISS appears to produce more ``false positives'' than PHOTO, these all arise in the dimmest magnitude bin $22-23$ where more ambiguity is present.
In this scenario, mistakes in COADD (our imperfect proxy of ground truth) likely favor PHOTO as both COADD and PHOTO are both produced by the same software pipeline.
We found several cases where BLISS detected a dim object that was not present in COADD, but confirmed its existence in the DECaLS survey.
\cref{fig:borderline}b is one example of a COADD mistake.
More importantly, BLISS correctly identifies the ambiguity in these situations, as most of the ``false positives'' BLISS finds have probabilities close to the detection threshold.
Setting this detection threshold allows practitioners to decide the level of certainty they require in sources for use in downstream tasks.


\section{Conclusion}
\label{conclusion}

BLISS is a fundamentally different approach to interpreting astronomical images. It uses deep learning to enable scalable and accurate Bayesian inference. BLISS performs well at detecting and deblending light sources in SDSS images.
BLISS is also highly extensible in that its inference routine requires little modification if the underlying statistical model is revised or extended.
We encourage interested researchers to try our software, which is available from \url{https://github.com/prob-ml/bliss}.

\clearpage
\bibliography{references}
\bibliographystyle{icml2022}





\end{document}